\documentclass[12pt]{iopart}
\usepackage{graphicx}
\newcommand{\g}{$\gamma$}
\newcommand{\eps}{$\epsilon(\theta)$}
\newcommand{\ff}{{\it ff}}
\usepackage[square,sort&compress,numbers]{natbib}
\usepackage[percent]{overpic}

\usepackage{color}
\usepackage{xspace}
\usepackage[normalem]{ulem}

\begin{document}

\title[Anisotropy in YBCO from Microwave Measurements]{Intrinsic Anisotropy and Pinning Anisotropy in Nanostructured YBa$_2$Cu$_3$O$_{7-\delta}$ from Microwave Measurements}

\author{N. Pompeo$^1$, A Alimenti$^1$, K Torokhtii$^1$, E Bartolom\'e$^2$, A Palau$^3$, T Puig$^3$, A Augieri$^4$, V Galluzzi$^4$, A Mancini$^4$, G Celentano$^4$, X Obradors$^3$, E Silva$^1$}

\address{$^1$ Dipartimento di Ingegneria, Universit\`a Roma Tre, Via Vito Volterra 62, 00146 Roma, Italy}
\address{$^2$ Escola Universitaria Salesiana de Sarri\`a (EUSS), Passeig Sant Joan Bosco 74, B08017-Barcelona, Spain}
\address{$^3$ Institut de Ci\`encia de Materials de Barcelona (ICMAB) - CSIC, Campus UAB, 08193-Bellaterra, Spain}
\address{$^4$ ENEA-Frascati Research Centre, 00044 Frascati, Italy}

\ead{enrico.silva@uniroma3.it}
\vspace{10pt}
\begin{indented}
\item[] January 28th, 2020
\end{indented}

\begin{abstract}
Anisotropy is an intrinsic factor that dictates the magnetic properties of YBa$_2$Cu$_3$O$_{7-\delta}$, thus with great impact for many applications. Artificial pinning centres are often introduced in an attempt to mitigate its effect, resulting in less anisotropic electrical and magnetic properties. However, the nanoengineering of the superconductor makes the quantification of the anisotropy itself uncertain: the intrinsic anisotropy due to the layered structure, quantified by the anisotropy factor $\gamma$, mixes up with the additional anisotropy due to pinning. As a consequence, there is no consensus on the experimental anisotropy factor $\gamma$ that can result in YBa$_2$Cu$_3$O$_{7-\delta}$ when directional  (twin planes, nanorods) or isotropic defects are present. We present here measurements of the magnetic field and angular dependent surface impedance in very different nanostructured YBa$_2$Cu$_3$O$_{7-\delta}$ films, grown by chemical route and by pulsed laser deposition, with different kind of defects (nanorods, twin planes, nanoparticles). We show that the surface impedance measurements are able to disentangle the intrinsic anisotropy from the directional pinning anisotropy, thanks to the possibility to extract the true anisotropic flux--flow resistivity and by correctly exploiting the angular scaling. We find in all films that the intrinsic anisotropy $\gamma = 5.3\pm0.7$. By contrast, the pinning anisotropy determines a much complex, feature--rich and nonuniversal, sample--dependent angular landscape.
\end{abstract}

%
%
%
%
%

\section{Introduction}
\label{sec:intro}

Anisotropy is a key issue in every theoretical treatment, data interpretation and technological application of cuprate superconductors. Whether it is a fascinating field of study \cite{Jin1989,Naughton1988,Kobayashi1995,Roulin1996,Blatter1992,Blatter1994}, an undesired feature for applications \cite{Larbalestier2001} or a property that can be tailored for specific purposes \cite{Maiorov2009,Sieger2017}, anisotropy requires to be dealt with. In YBa$_2$Cu$_3$O$_{7-\delta}$ (YBCO), the cuprate superconductor we  focus on in this paper, the matter is made particularly rich by the great relevance gained by nanostructuring in order to increase the performances for high--current applications \cite{Gutierrez2007a,Foltyn2007}. In fact, improving the current--carrying performances requires hindering fluxon motion, thus introducing defects of various dimensionality. Elongated defects (1D) like nanorods are typical of materials grown by Pulsed Laser Deposition (PLD) \cite{Goyal2005}, while large (on the scale of the coherence length) nanoparticles constitute 3D defects \cite{Cayado2015}, typical of chemical methods for the growth. In addition, twin planes and grain boundaries are essentially ubiquitous and represent planar (2D) defects. Point--like defects, of size smaller than the coherence length are 0D defects. All of the defects listed can coexist, or can be made to coexist \cite{Foltyn2007,Maiorov2009,Sieger2017} to improve the overall performances.

YBCO, in its turn, is an intrinsically anisotropic superconductor, due to its well--known layered structure. This feature brings in an anisotropic electron effective mass \cite{Campbell1988}, that for the present purposes can be treated as uniaxially anisotropic, with $m_c=\gamma ^2m_{ab}$, where $m_c, \; m_{ab}$ are the anisotropic masses along the $c$ axis and $ab$ planes, respectively, and \g\; is the anisotropy coefficient that we will indicate here and henceforth as {\em intrinsic anisotropy} as opposed to the anisotropy due to defects, that we will indicate as {\em pinning anisotropy} \footnote{For completeness, it should be mentioned that the layered structure itself can act as a pinning anisotropy when an external field is very close to the alignment with the $ab$ planes \cite{Doyle1993,Kwok1991}. This kind of anisotropic pinning is usually referred to as ``intrinsic pinning", which unfortunately does not add to the clarity of the nomenclature.}.

Thus, two sources of anisotropic behaviour exist: the intrinsic anisotropy, and the pinning anisotropy. Broadly speaking, thermodynamic properties are mostly affected by the intrinsic anisotropy, while pinning--related properties have a prominent contribution from the anisotropy introduced by defects. 
For what concerns transport properties, it was argued that the free--flux--flow resistivity is affected by the anisotropy in the same fashion as thermodynamic quantities \cite{Hao1992}. In fact, the flux--flow resistivity is an intrinsic property and it is not in principle affected by pinning, unless the introduction of defects significantly changes some basic electronic properties (e.g., the quasiparticle scattering rate). In the latter case the material changes substantially, with consequences on the superconducting quantities such as the coherence length and the upper critical field. In practice, the dc flux--flow resistivity can be measured only very close to the $H_{c2}(T)$ transition, not just but well above the irreversibility field, so that the investigation has been limited to the transition region.
In such a complex scenario, it is not surprising that there is still not a general consensus on the value of the anisotropy ratio $\gamma$ in nanostructured YBCO, due to the difficulty to identify from the experiments the different contributions. 

Before recalling some of the vast literature on the matter, it is necessary to mention how a measure of the anisotropy is usually given. As a reminder on notation, we recall that:
\begin{equation}
\label{eq:gamma}
\gamma=\sqrt{m_c/m_{ab}}=\xi_{\parallel}/\xi_{\perp}=H_{c2\parallel}/H_{c2\perp},
\end{equation}
where the subscript $\parallel$ ($\perp$) indicates parallel (perpendicular) to the layers, and $\xi$ and $H_{c2}$ are the coherence lengths and the upper critical fields, respectively.

\begin{figure}
\includegraphics[width=8 cm]{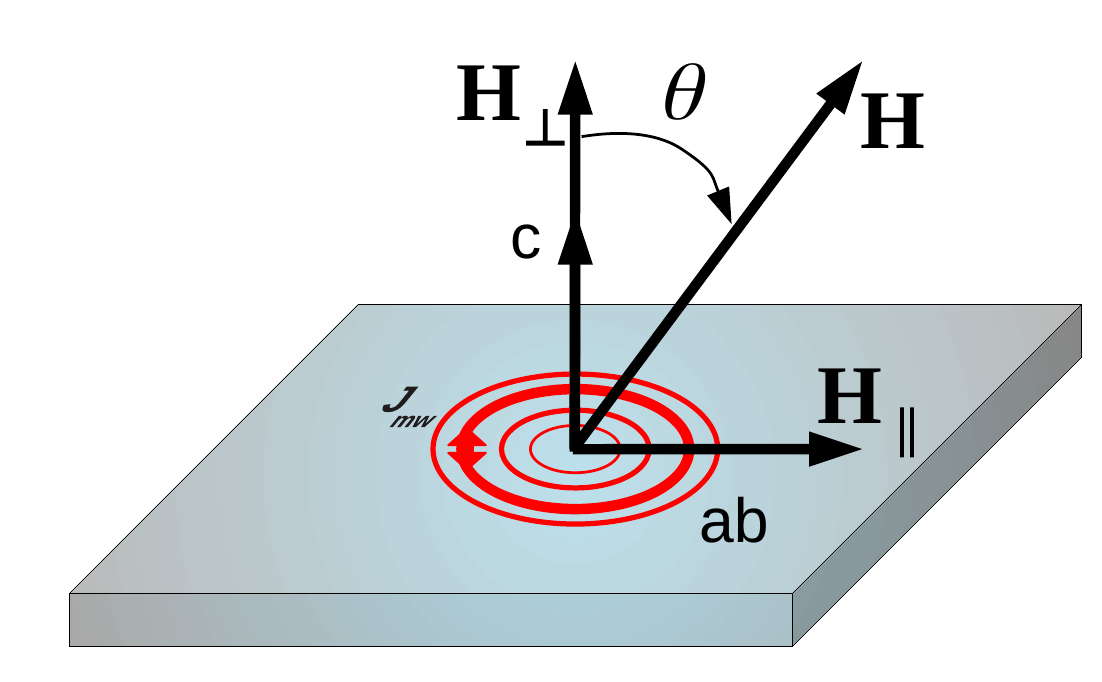}\\
\caption{Sketch of the geometrical configuration and definition of the angles. $\theta$ is defined as the angle between the applied field $H$ and the $c$ axis. $\perp$ and $\parallel$ indicate the alignment with the $c$ axis and the $ab$ planes, respectively. The lines of the microwave current density (see Sec.\ref{sec:exp}) are also depicted.}
\label{fig:geometry}
\end{figure}

In the early measurements on synthetic multilayered materials \cite{Jin1989,Nakajima1989} the anisotropy was determined by simply measuring $H_{c2//}$ and $H_{c2\perp}$, whence \g. In some cases, the measurements where taken at different angles $\theta$ with respect to the $c$ axis (Figure \ref{fig:geometry} presents a sketch of the geometrical configuration), and compared to the anisotropic Ginzburg--Landau theory, which predicts \cite{Sudbo2004book}:
\begin{eqnarray}
H_{c2}(\theta)=
\frac{H_{c2,\perp}}{\left(\gamma^{-2}\sin^2\theta+\cos^2\theta\right)^{1/2}}=H_{c2,\perp}/\epsilon(\theta)
\label{eq:Hc2}
\end{eqnarray}
where the second equality defines the \textit{anisotropy factor} \eps. This procedure is not easily feasible in YBCO: not only $H_{c2}$ attains huge values but, in addition, very close to $T_c$, where $H_{c2}$ becomes sufficiently small, very strong thermal fluctuations make an accurate experimental definition of the upper critical field very difficult. However, in  independent papers \cite{Blatter1992,Hao1992,Hao1993,Blatter1993}, it was shown that 
in high--$\kappa$ superconductors and 
under appropriate circumstances every angle and magnetic--field--dependent observable $Q(H,\theta)$ depended on the \textit{rescaled field} $H/H_{c2}(\theta)\propto H\epsilon(\theta)$ only, as:%
\begin{eqnarray}
Q(H,\theta)=\alpha(\theta)Q(H\epsilon(\theta))
\label{eq:scaling}
\end{eqnarray}
where $\alpha(\theta)$ depends on the observable \cite{Blatter1992,Hao1992,Blatter1994} (for the cases here relevant, $\alpha(\theta)=1$). This property was named ``BGL scaling rule". It allowed to evaluate \g\, by measuring an observable at different fields and angles, and looking for the collapse of different curves when plotted against a suitable $H\epsilon(\theta)$, whence an experimental \eps\, and ultimately (through a fit of the angular function \eps\, by, e.g., Eq.\ref{eq:Hc2}), \g. It must be emphasized further that the scaling rule takes place for observables where only the intrinsic anisotropy is responsible for the anisotropic behaviour. As a consequence, all 1D, 2D, 3D defects\footnote{0D (point--like) defects do not add any anisotropy, but 3D defects do: a 3D (e.g., spherical) object is a perturbation to an anisotropic matrix.} can mask the resulting intrinsic \g\, or the same scaling rule can break down.
\footnote{
A more subtle origin for the breaking of the scaling is a field and angular dependent microscopic basis of superconductivity: in two--band superconductors, the presence of a weaker band and the effect of interband scattering on $T_c$ may affect the persistence of the angular scaling.
}
%
The first studies in pure, pristine YBCO, in the shape of single crystals, reported $\gamma \simeq 7$ by means of different techniques: specific heat \cite{Roulin1996}, resistivity \cite{Naughton1988}, or magnetization \cite{Kobayashi1995}. Notably, the measurements of the resistivity were used to evaluate the upper critical field in the high part of the resistive transition, since at low resistance levels the fluxon dynamics is strongly affected by pinning.  Thin films, when pristine, gave similar results \cite{Sarti1997a,Sarti1997b}: the scaling procedure could be performed again on the upper part of the transition, where only thermal fluctuations or possibly free--flux flow contributed to the dissipation. By contrast, the evaluation of the anisotropy from transport properties in nanostructured YBCO films is highly diversified \cite{Civale2004,Traito2006,Gutierrez2007a}, although comparison of different methods helped to shed light on the matter \cite{Bartolome2019}.
In fact, it was reported that the addition of nanodefects to the YBCO matrix could affect the upper critical field \cite{Gutierrez2007a,Malmivirta2014,Xu2015b,Malmivirta2016}, by the nanodefects themselves or by the resulting strain. Since $H_{c2}$ derives from the coherence length (affected by the mean free path) and from the effective mass (the band structure), see Eq.\ref{eq:gamma}, it is possible that the addition of defects has some influence on $H_{c2}$, and then on the anisotropy itself as determined from the scaling rule. We underline that this effect, if present, is independent on the pinning that can be produced: as an example, a negligible effect on the anisotropy, indicating negligible change in band structure, can however be accompanied by a strong directional pinning, like it was found in YBCO with BZO nanorods \cite{Pompeo2013} and with nanoparticles \cite{Bartolome2019}. However, in some films grown by PLD with different nanoparticles, a significant reduction of the anisotropy was reported \cite{Palonen2013,Malmivirta2014,Malmivirta2016}, accompanied by an increase of $H_{c2}$, although in different films the increase of $H_{c2}$ did not affect the anisotropy \cite{Xu2015b}. It is then interesting to investigate the anisotropy of YBCO with different nanostructuring, such as nanoparticles and nanorods.

The aim of this paper is to present an experimental method based on the measurement of the field--dependent microwave surface impedance in thin YBCO films, able to discriminate between intrinsic and pinning anisotropy. We exploit the capabilities of the high--frequency measurements to yield information about the free--flux--flow resistivity and the pinning efficiency through the so--called depinning frequency from the same set of measurements. We perform measurements on very different nanostructured YBCO thin films: pristine and nanostructured, with artificial pinning centres in the form of nanorods or nanoparticles. We exploit the scaling procedure of the free--flux--flow resistivity to obtain the anisotropy  coefficient \g\; and we obtain, in all samples, \g$=5.3\pm0.7$, in agreement with the multi-technique investigation in \cite{Bartolome2019}, despite the very different defect anisotropy, covering here 1D (nanorods), 2D (twin planes) and 3D (nanoparticles) defects. The remarkably diverse pinning anisotropy is however evident in the depinning frequency. 

The paper is organised as follows: in Sec.\ref{sec:micro} we summarize the main features of the fluxon motion physics in the high--frequency regime, including the peculiarities of the anisotropic measurements. In Sec.\ref{sec:exp} we describe the experimental setup and the measurement chain, as well as the samples under study. In Sec.\ref{sec:scaling} we report on the measurements of the flux--flow resistivity, the angular scaling and the results for the anisotropy factor \g. In Sec.\ref{sec:pinning} we discuss the data for pinning anisotropy. We shortly conclude in Sec.\ref{sec:conc}.

\section{Flux motion at microwave frequencies}
\label{sec:micro}
For the purposes of this paper, the peculiarities of the response of the fluxon system to a high--frequency alternating current are the key factors that allow to ultimately determine the intrinsic anisotropy \g. We then briefly recall here the physics of the high--frequency vortex motion.

As it is known \cite{Tinkham1996book}, a current density $J$ exerts on a flux quantum $\Phi_0$ the Lorentz force (per unit length) $F_L=J \times \Phi_0$. At sufficiently high frequency, and low driving amplitude (to be specified later in Sec.\ref{sec:exp}) the alternating force determines only very small oscillations of the flux quanta around their equilibrium positions, of the order of a fraction of a nm \cite{Tomasch1988a}. Although the equilibrium position of the flux quanta results from the competition between pinning centres and vortex--vortex interaction, at sufficiently high frequencies such equilibrium configuration is not perturbed by the stimulus. This fact has as a direct consequence that a single--vortex equation of motion can be applied, and the dynamic problem greatly simplifies \cite{Gittleman1966,Pompeo2008} \footnote{When $J$ is a dc current the framework is totally different, no vortex motion exists until the Lorentz force exceeds the pinning force, and the motion that results can be severely affected by the vortex configuration: many different vortex phases can exist, each of which with a different dynamic regime until only at temperatures and fields very close to the transition the free--flux--flow regime is eventually reached.}. It is then appropriate to write down the single--vortex equation of motion, and interpret the resulting vortex parameters as average (over the flux quanta whose motion is probed) values. The equation of motion of a vortex subjected to the Lorentz force $F_L$, displaced by $x$ from its equilibrium position, with velocity $v$, reads:
\begin{eqnarray}
k_px+\eta v= F_L+F_{th}
\label{eq:force}
\end{eqnarray}
where the smallness of the vortex displacement allowed the elastic approximation for the recall force $k_p x$, $k_p$ is the pinning constant or Labusch parameter \cite{Beasley1969}, $\eta$ is the vortex viscosity (or drag coefficient) \cite{Bardeen1965}, responsible for the power dissipated by a moving vortex, and $F_{th}$ represents the thermal forces acting on the vortex and ultimately responsible for the flux--creep or other thermally activated phenomena. The inertial term has been neglected as customary \cite{Golosovsky1996}, due to the smallness of the estimated vortex mass \cite{Blatter1994,Golosovsky1996}. Equation \ref{eq:force} gives rise to the following result for the complex vortex--motion ac resistivity \cite{Gittleman1966,Brandt1992,Coffey1991a,Pompeo2008}:
\begin{eqnarray}
\rho_{v}=\rho_{v1}+\rm{i}\rho_{v2}=\rho_{\ff}\frac{\chi+\rm{i}\frac{\nu}{\bar{\nu}}}{1+\rm{i}\frac{\nu}{\bar{\nu}}}
\label{eq:complexrho}
\end{eqnarray}
where $B\simeq\mu_0 H$ (London approximation), $\rho_{\ff}=\Phi_0 B/\eta$ is the true flux--flow resistivity \footnote{We stress that this value equals the flux--flow resistivity which would appear if only Lorentz force and viscous drag were present.}, $\chi \in [0,1]$ is a creep factor ($\chi=0$, no creep, $\chi = 1$, maximum creep, all fluxons depinned), and $\bar{\nu}$ is a crossover frequency between a pinning--unsensitive response, $\nu \gg \bar{\nu}$, where $\rho_v\simeq\rho_{\ff}$, and the regime  $\nu \ll \bar{\nu}$ where, depending on the weight of the thermal forces and hence the creep factor, the response can be mostly reactive and pinning--dominated ($\chi\ll 1$) or dictated by thermal forces.

A discussion of the many different models leading to Equation \ref{eq:complexrho}, as well as different regimes, has been given elsewhere \cite{Pompeo2008,Silva2017book}, and we address to these earlier publications for a thorough discussion. Here we want to stress the following feature that is essential to the data analysis: due to a combination of analytical and physical constraints, it can be shown that from the experimental parameter $r=\rho_{v2}/ \rho_{v1}$ one can evaluate the maximum value for $\chi$ compatible with the data, $\chi_M=1+2r^2-2r\sqrt{1+r^2}$ \cite{Pompeo2008}. Then, it is possible to obtain uncertainty bars on the values of $\rho_{\ff}$ and $k_p$ on the basis of $\chi_M$. For not too large $\chi_M < 0.5$ it is safe (to approximately 20\%) to rely on the simplified equation for the complex resistivity derived in the early days of the studies of the high--frequency study of vortex motion \cite{Gittleman1966}:
\begin{eqnarray}
\rho_{v}=\rho_{v1}+\rm{i}\rho_{v2}=\rho_{\ff}\frac{1}{1-\rm{i}\frac{\nu_p}{\nu}}
\label{eq:complexrhogr}
\end{eqnarray}
where now $\nu_p=k_p/2\pi\eta$ is the so-called {\em depinning frequency}, an important parameter in high--frequency applications \cite{Calatroni2017b} and, as we show in the following, a useful quantity in the study of the anisotropy.

Summarizing, from the complex vortex motion resistivity one can reliably extract \cite{Pompeo2008} the depinning frequency $\nu_p$ and the flux--flow resistivity $\rho_{\ff}$, and derive the pinning constant $k_p$. It is now important to stress that the flux--flow resistivity is an observable related to the fundamental processes of Cooper pairs -- quasiparticle conversion and of quasiparticle scattering \cite{Tinkham1996book}. Thus, it is a quantity tightly bound to the superconducting and normal state and almost unrelated to pinning, unless defects become so prominent as to influence the normal state scattering, thus changing the very nature of the material -- YBCO in this case.
In the latter case the basic superconducting quantities, such as the coherence length and the critical fields are expected to change.
By contrast, $k_p$ is directly related to the strength and nature of pinning centres only. It now appears how high--frequency measurements can yield separate information on fundamental and defect--related properties.

Once the nature of the information that can be gained from microwave measurements is settled, we move to the role of anisotropy. Equation \ref{eq:force} is in fact a vector relation, and allowing for arbitrary field and current direction it becomes a tensorial problem. When both angles that the magnetic field makes with the $c$ axis and with the driving current density $J$ change in an arbitrary way, in an anisotropic medium the motion of the flux lines is no more perpendicular to the plane identified by $H$ and $J$ and this effect leads to a measured resistivity that cannot be straightforwardly identified with some $ab$ plane or $c$ axis components \cite{Pompeo2015}. It is then necessary to specify the measuring geometry. In our setup (Sec.\ref{sec:exp}), we have circular--symmetric microwave currents applied along the sample $ab$ planes, as depicted in Fig.\ref{fig:geometry}. This feature implies that the Lorentz force changes with varying $\theta$. The resulting electromagnetic problem has been solved earlier \cite{Pompeo2015,Pompeo2018a}. For the purpose of the analysis of the data, the essential features are as follows: 
\begin{itemize}
\item{the measured vortex--motion in--plane resistivity $\rho_{vm}$ does not in general coincide with the material property $\rho_v$, essentially due to the spatially varying Lorentz force giving rise to an additional angular dependence. Accordingly, the measured flux--flow resistivity $\rho_{ff m}$ differs from the material property $\rho_{ff}$ \cite{Pompeo2015};}
\item{the depinning frequency $(\nu_p)$ has the very remarkable property of being a scalar quantity when only point pins are present (unlike $\rho_{\ff}$ and $k_p$, which are tensors). Thus, the measured depinning frequency coincides with the the sample--specific corresponding property (sample--specific because $\nu_p$ includes the effect of pinning) \cite{Pompeo2018a};}
\item{the depinning frequency $\nu_p(H,\theta)$, in the case of point--pinning, is expected to follow the scaling rule \cite{Pompeo2018a}. Any deviation from a scaling behaviour is a direct demonstration of the existence of directional pinning.}
\end{itemize}
The resulting relations come out to be \cite{Pompeo2015,Pompeo2018a}: 
\begin{eqnarray}
\label{eq:rhoff}
  \rho_{\ff m}(H,\theta) =\rho_{\ff}(H,\theta) f_{L}(\theta) \\
\label{eq:nup}
  \nu_{p,meas}(H,\theta) =\nu_p(H,\theta) =\nu_p(H\epsilon(\theta)) \\
\label{eq:fteta}
  f_{L}(\theta)=\frac{\frac{1}{2}\gamma^{-2}\sin^2\theta+\cos^2\theta}{\gamma^{-2}\sin^2\theta+\cos^2\theta}
\end{eqnarray}
where $f_L(\theta)$ is the correction to the angular dependence in our experimental geometry (general expressions of $f_L$ were worked out in \cite{Pompeo2015,Pompeo2018a}). These equations will prove useful in the interpretation of the data and in the extraction of \g.

\section{Experimental setup and samples}
\label{sec:exp}
The experiments were performed on different nanostructured YBCO thin films. We have chosen on purpose films with very different nanostructure, in order to assess the question of whether different nanostructuring can change the intrinsic anisotropy. We performed extended measurements on a set of CSD (Chemical Solution Deposition method) epitaxial films (pristine and with nanoparticles) and we compared the results to measurements taken in PLD films with BaZrO$_3$ (BZO) nanorods.  All samples had $T_c\simeq$ 90 K.

CSD epitaxial films were grown at ICMAB-CSIC \cite{Llordes2012} on 5$\times$5 mm$^2$ LaAlO$_3$ substrates using a metal-organic decomposition method based on the TFA route \cite{Xu2015,Llordes2012}. We focus here on three specific samples: CSD1 is a pristine YBCO sample, CSD2 is a 6\%--added Ba$_2$YTaO$_6$ (BYTO) film and CSD3 is a 12\%--added BaHfO$_3$ (BHO) film grown by a flash heating method \cite{Li2019}. 
A detailed description of all kinds of CSD films 
here examined
can be found elsewhere \cite{Coll2013a, Bartolome2019}. In all CSD nanostructured films 3D defects were present. The vortex pinning mechanism in these nanocomposites has been shown to be mostly dominated by the presence of highly strained areas in the YBCO matrix, i.e. nanostrain ($\varepsilon$). Twin planes and grain boundaries were also present in all films to a variable extent. Pristine CSD1 has a thickness $t_s=$ 180 nm, nanostrain of $\varepsilon$ = 0.12\% and self-field $J_c$ at 77 K $J_c$ = 3.4 MA$\cdot$cm$^{-2}$. Nanocomposites CSD 2 and CSD 3, presented values of $t_s=$200 nm, $\varepsilon$ = 0.16\%, $J_c$ = 5.1 MA$\cdot$cm$^{-2}$ and $t_s=$ 150 nm, $\varepsilon$ = 0.25\%, $J_c$ = 2.4 MA$\cdot$cm$^{-2}$, respectively.
\begin{figure}[ht]
\hspace{1cm}
\begin{overpic}[width=6 cm]{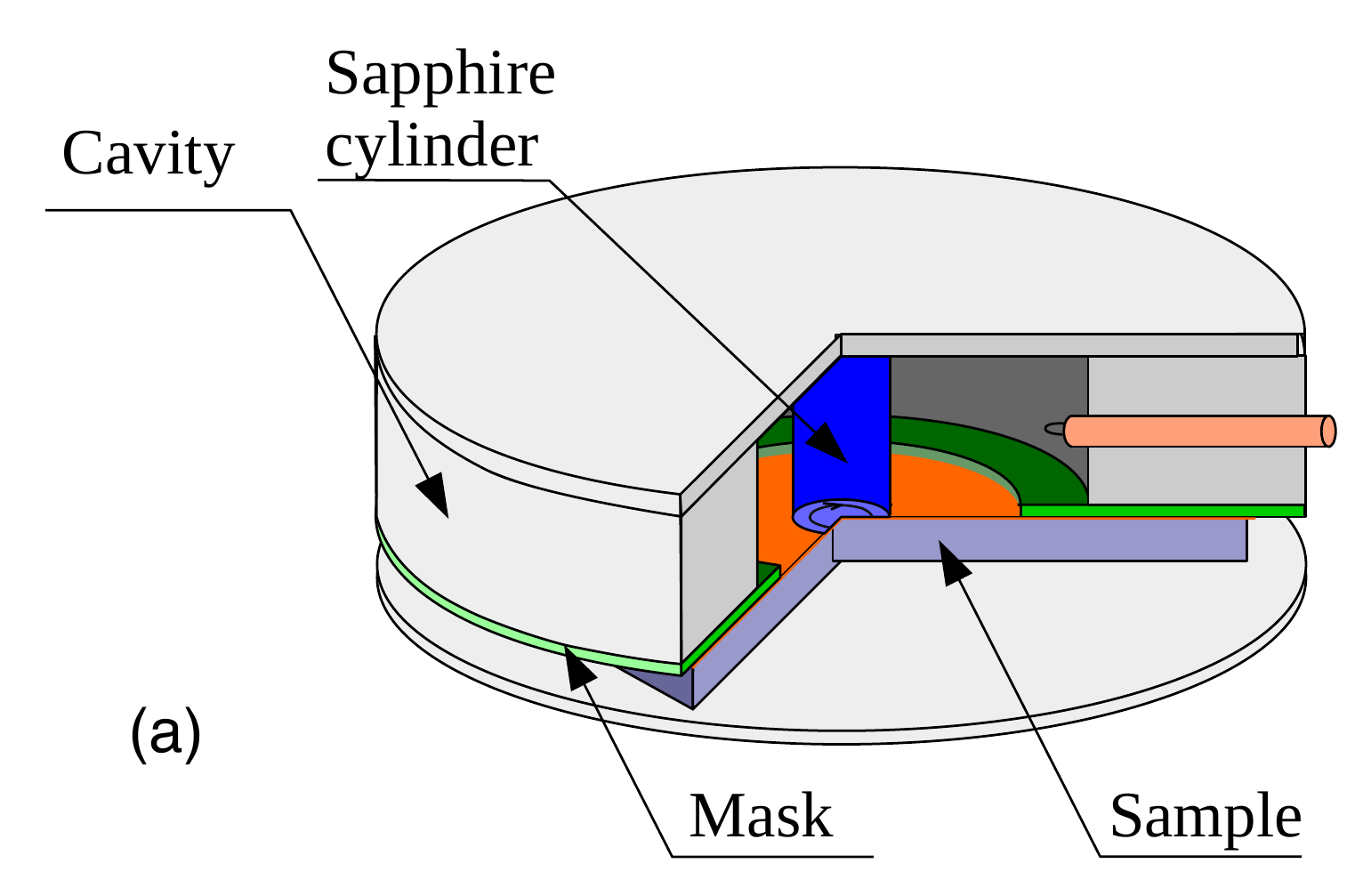}\put (10,10) {(a)}\end{overpic}\\
\begin{overpic}[width=8 cm]{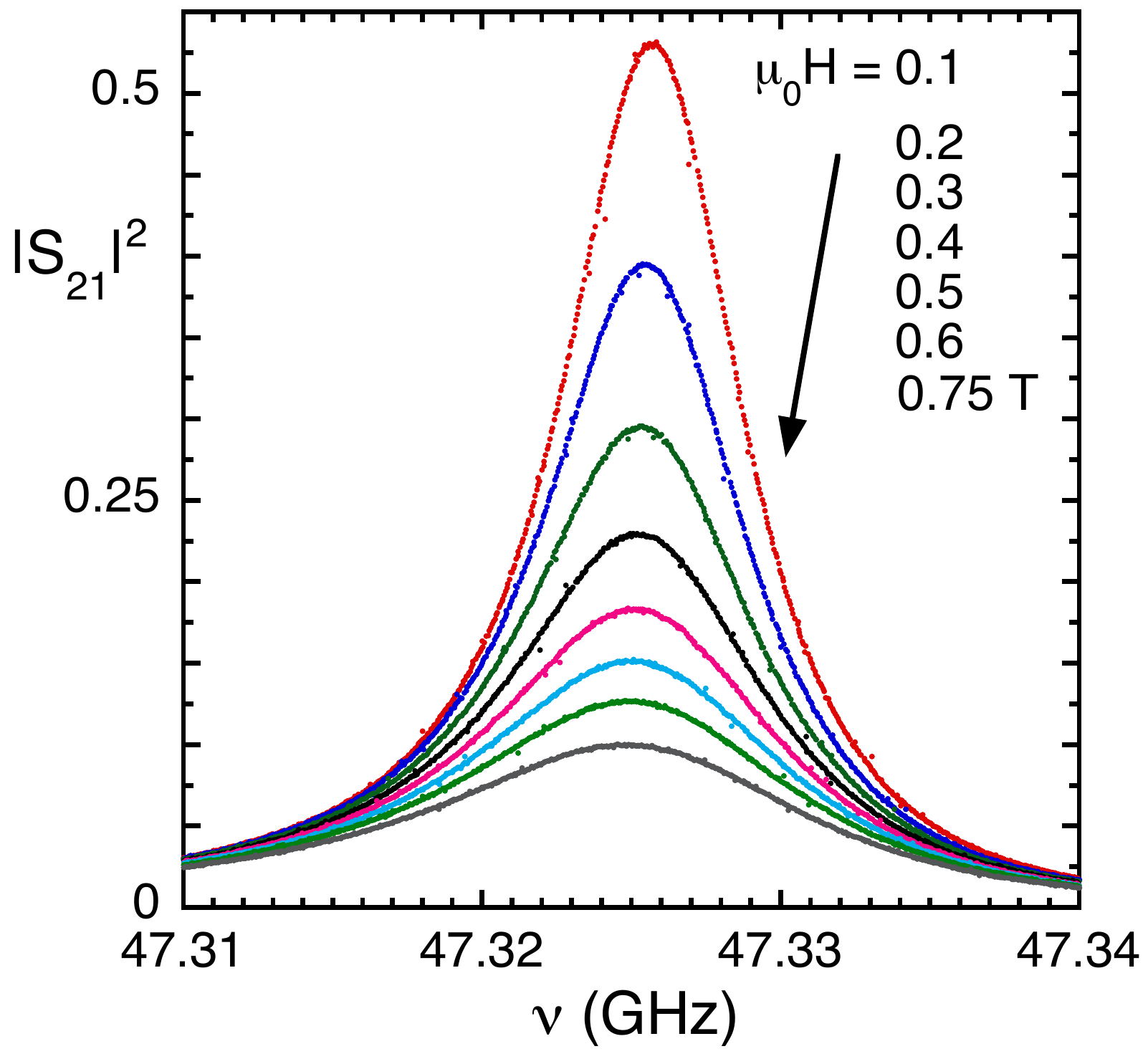}\put (21.5,80) {(b)}\end{overpic}
\caption{(a) Sketch of the cylindrical dielectric--loaded resonator (straight section, not to scale). The superconducting film replaces one base of the resonator. The film edges are covered by a metal mask. One of the coupling loops is also depicted. (b) Change of the transmitted signal $|S_{21}|$ as a function of the field at $T$ = 83 K: the shift of the resonance frequency $\nu_0$ and the broadening of the curve, indicating a decrease of $Q$ with the field, are clearly visible.}
\label{fig:exp}
\end{figure}

The PLD YBCO films were grown at ENEA--Frascati on 7.5$\times$7.5 mm$^2$ SrTiO$_3$ substrates from targets with BZO powders at 5\% mol. (samples PLD1 and PLD2) and 7\% mol. (sample PLD3) \cite{Galluzzi2007}, with a resulting thickness of $\sim$120 nm. BZO promoted the formation of columnar-like defects (nanorods), approximately perpendicular to the film plane, as seen from transverse TEM images \cite{Augieri2010}. Such nanorods add to the existing twin planes and grain boundaries. The density of columns corresponds to an equivalent matching field of about 1.2 T in PLD1 and PLD2, and 2 T in PLD3. Separate dc investigations at $T=$ 77 K reported self-field $J_c\simeq$3.7~MA$\cdot$cm$^{-2}$ in samples PLD1 and PLD2 and 1.6~MA$\cdot$cm$^{-2}$ in sample PLD3.

The response function at  microwave frequencies is the so--called surface impedance $Z_s=R_s+{\rm i}X_s=E_{\parallel}/H_{\parallel}=\sqrt{\rm{i}\mu_0 2\pi\nu\rho}$, where $E_{\parallel}$ ($H_{\parallel}$) is the electric (magnetic) field parallel to the surface of the superconducting film, $\nu$ is the frequency of the microwave fields, and $\rho$ is the complex resistivity which includes contributions from the vortex motion as well as from superfluid and quasiparticles.

The field-- and angular-- dependent surface impedance was measured by the well--known \cite{Chen2004,Pompeo2014,Pompeo2017} dielectric resonator method. In our case, we use a cylindrical copper cavity loaded with a cylindrical sapphire puck \cite{Pompeo2014,Alimenti2019a}. The structure has a resonance frequency at $\nu_0\sim\;$47.5 GHz, thus at the high frequency edge of the microwave range. This high operating frequency ensures the reliability of the single--vortex equation which is at the ground of the model used to analyse the data: at this frequency and with an estimated microwave power $<$ 0.1 mW in the cavity, following \cite{Tomasch1988a} we estimate a vortex displacement of less than 0.1 nm. Figure \ref{fig:exp}a reports a sketch of the resonator and of the placement of the superconducting film.

We apply a moderate field $\mu_0H < $~0.8~T at different field orientations $\theta$, and we record at fixed temperature ($T\geq$~63 K) the change in the resonance shape with the magnetic field and angle. The temperature is held within $\pm\, 10$ mK at most during each set of measurements. By appropriate fits of the resonance shape, including non--idealities \cite{Alimenti2019a,Pompeo2017a}, we obtain the unloaded quality factor $Q$ and the resonance frequency $\nu_0$. The changes in $Q$ and $\nu_0$ yield the changes in the surface impedance $\Delta Z_s$ according to \cite{Pompeo2014}:
\begin{eqnarray}
\label{eq:Zsmeas}
\nonumber
\Delta Z_s(H,\theta)=\\
=G_s\left[\frac{1}{Q(H,\theta)}-\frac{1}{Q(H=0)}\right]
-2{\rm i}G_s\left[\frac{\nu_0(H,\theta)-\nu_0(H=0)}{\nu_0(H=0)}\right]
\end{eqnarray}
where the geometrical factor $G_s$ is calculated analytically and checked with electromagnetic simulations. Since (i) we are dealing with thin films, of thickness $t_s  < 2\lambda$ ($\lambda$ is the temperature--dependent London penetration depth), and (ii) in moderate fields the change in superfluid and quasiparticles is negligible with respect to the vortex motion, one can write to an approximation better than 5\% \cite{Pompeo2017b}:
\begin{eqnarray}
\label{eq:thin}
\Delta Z_s(H,\theta)\simeq \frac{\Delta \rho_{vm}(H,\theta)}{t_s}=  \frac{\rho_{vm1}(H,\theta)+{\rm i} \rho_{vm2}(H,\theta)}{t_s}
\end{eqnarray}
By putting together Eq.s \ref{eq:Zsmeas} and \ref{eq:thin} it is straightforward to conclude that the vortex motion complex resistivity is directly accessible from the experimentally measured $Q$ and $\nu_0$. Figure \ref{fig:exp}b reports the change in the resonance curve with increasing field, where the shift of $\nu_0$ and the decrease of $Q$ (broadening of the resonance curve) are apparent. 

\begin{figure}[ht]
\includegraphics[width=8 cm]{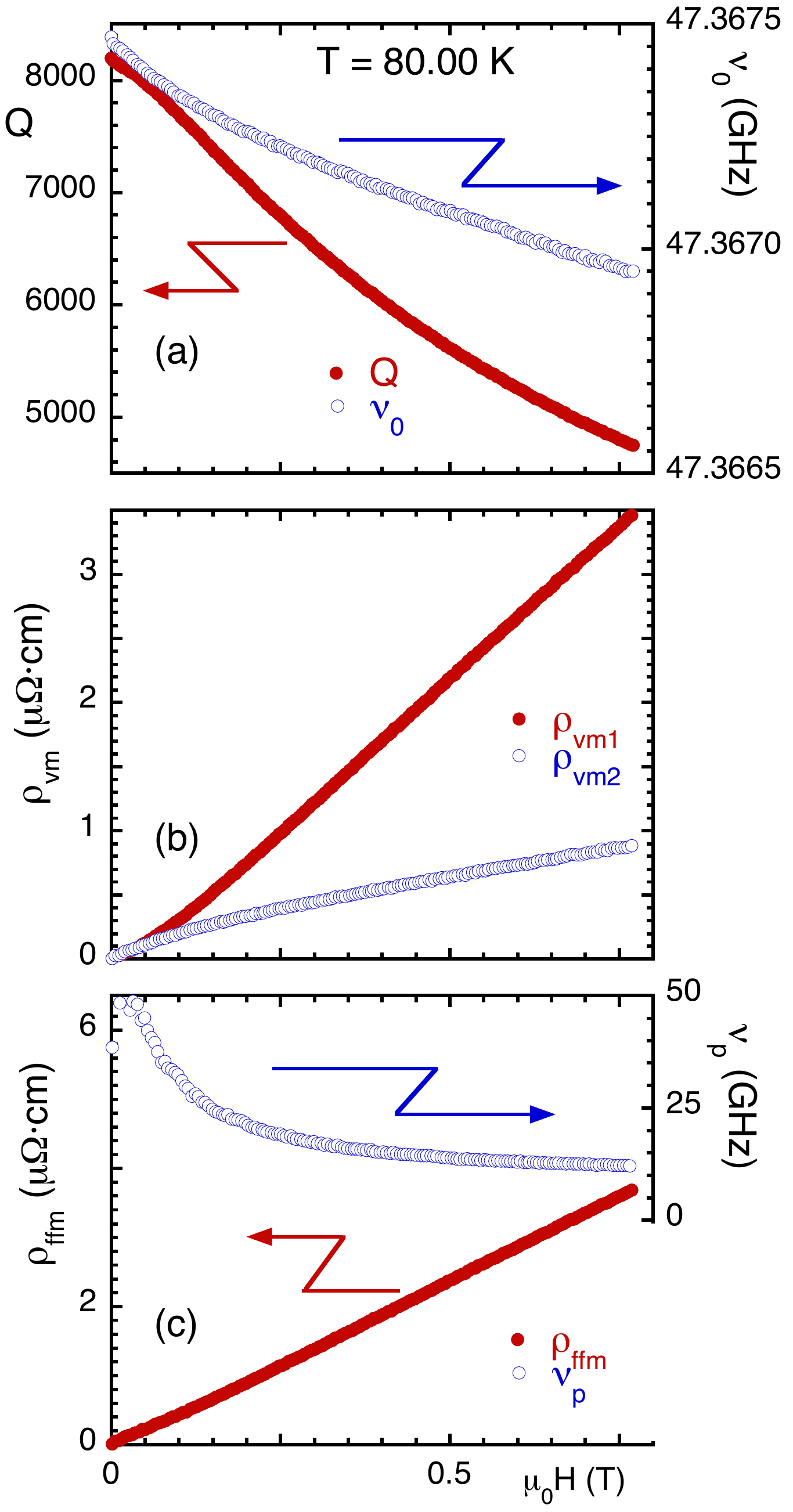}
\caption{Exemplification of the derivation of the vortex parameters as a function of the field $H$ from the raw data, at $T=80.00$~K and $\theta=0^{\circ}$: (a) raw data of the change of the quality factor $Q$ and of the resonance frequency $\nu_0$. (b) Corresponding change of the vortex motion resistivity, $\rho_{vm1}$ and $\rho_{vm2}$. (c) Data for the measured flux--flow resistivity, $\rho_{\ff m}$, and of the depinning frequency, $\nu_p$. Data taken on sample CSD1.}
\label{fig:measchainH}
\end{figure}

\begin{figure}[ht]
\includegraphics[width= 8cm]{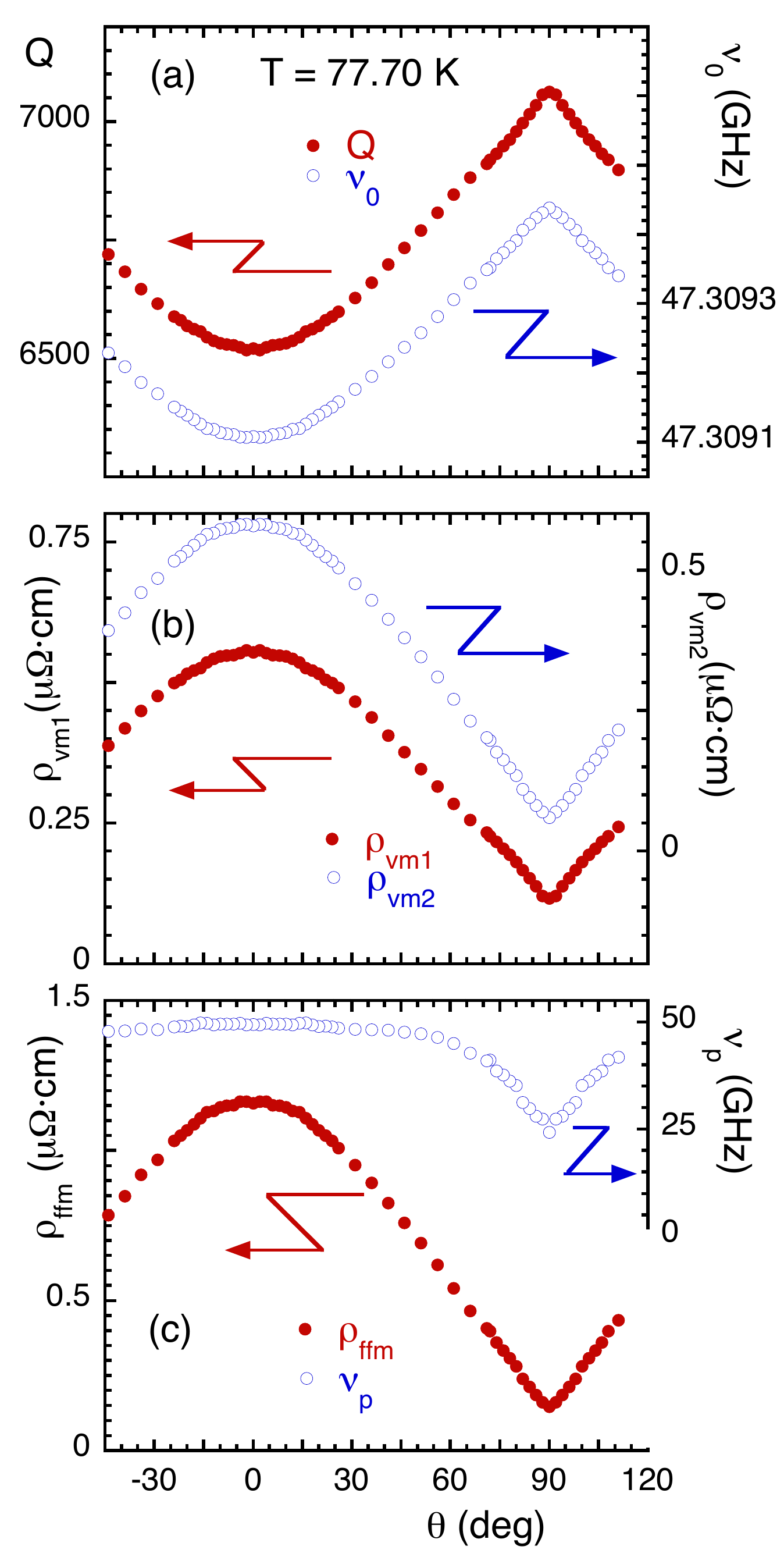}\\
\caption{Exemplification of the derivation of the vortex parameters as a function of the angle $\theta$ from the raw data, at $T=77.70$~K and $\mu_0H=0.4$~T: (a) raw data of the change of the quality factor $Q$ and of the resonance frequency $\nu_0$. (b) Corresponding change of the vortex motion resistivity, $\rho_{vm1}$ and $\rho_{vm2}$. (c) data for the measured flux--flow resistivity, $\rho_{\ff m}$, and of the depinning frequency, $\nu_p$. Data taken on sample CSD2.}
\label{fig:measchainteta}
\end{figure}

The complete measuring chain is exemplified in Figures \ref{fig:measchainH} and \ref{fig:measchainteta}. We show how we derive $\rho_{vm1}$ and  $\rho_{vm2}$ from the raw data for $Q,\,\nu_0$ on the basis of Eq.s \ref{eq:Zsmeas} and \ref{eq:thin}. Then, making use of the model of Eq.\ref{eq:complexrhogr}, we derive the depinning frequency $\nu_p$ and the flux--flow resistivity. For the sake of compactness of the notation, although the flux--flow resistivity is - strictly speaking - a derived quantity, we will call it ``measured flux--flow resistivity" and we will use the symbol $\rho_{ff m}$ as in Eq.\ref{eq:rhoff}. In the next Sections we  describe and analyse separately the scaling of the flux--flow resistivity and the properties of the depinning frequency in the different samples.

\section{Scaling of the anisotropic flux--flow resistivity}
\label{sec:scaling}
\begin{figure}[ht]
\includegraphics[width= 7cm]{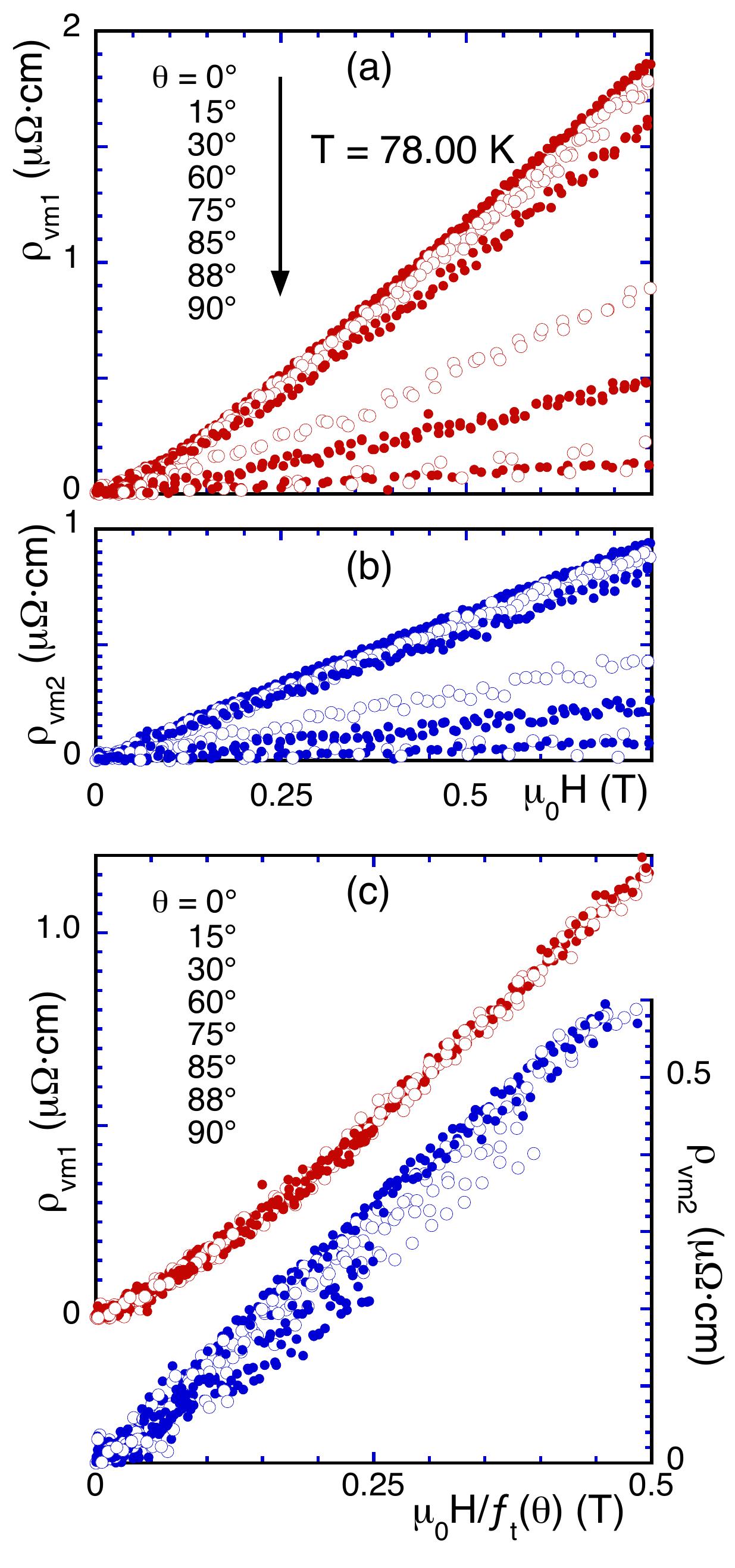}\\
\caption{Demonstration of the failure of the angular scaling for the full vortex motion resistivity: (a) measured real part of the vortex--motion resistivity $\rho_{vm1}(H)$ as a function of the applied field $H$ at selected different angles $\theta$. (b) Measured imaginary part of the vortex--motion resistivity $\rho_{vm2}(H)$ as a function of the applied field $H$ at different angles $\theta$. Full and open symbols are alternated to ease reading of the scaled data in the following panels. (c) Almost perfect scaling of the real part, $\rho_{vm1}(H/f_t(\theta))$ (red symbols) can be obtained by rescaling the field with a suitably chosen function $f_t(\theta)$, but the imaginary part, $\rho_{vm2}(H/f_t(\theta))$ (blue symbols) fails to scale with  the same rescaled field. Pinning prevents the angular scaling of a pinning--affected quantity. Data taken in sample CSD3, $T=$78 K. To avoid crowding, 30\% or less of the data is plotted.}
\label{fig:noscaling}
\end{figure}

We first examine the scalability of the full complex resistivity $\rho_{vm}$. This is not an obvious fact, as demonstrated by the following measurements and argument. In Figure \ref{fig:noscaling}a,b we report sample measurements of the \emph{complex} $\rho_{v}$, in terms of its real and imaginary parts, $\rho_{vm1}(H,\theta)$ and $\rho_{vm2}(H,\theta)$, taken at fixed angles and with sweeping the magnetic field. Figure \ref{fig:noscaling}c reports the same data scaled using  $\rho_{vm1}(H,\theta=0)$ as a master curve: we searched at each angle $\theta$ for a suitable factor $f_t(\theta)$ such that the curves $\rho_{vm1}(H/f_t(\theta))$ could be collapsed together: this is indeed possible, as shown in Figure \ref{fig:noscaling}c. However, when the curves for the imaginary part, $\rho_{vm2}(H,\theta)$, are plotted versus the same reduced field, $\rho_{vm2}(H/f_t(\theta))$, they {\em do not scale}. It is also possible to perform the reverse procedure (not shown): one can scale $\rho_{vm2}$, thus obtaining a different $H/f_t(\theta)$, but then $\rho_{vm1}$ does not scale with the new $H/f_t(\theta)$. That is, the angular scaling of the full complex resistivity does not take place and it cannot be used to determine $\gamma$. This is a direct demonstration that in pinning--affected quantities (such as $\rho_{vm}$, see Eq.\ref{eq:complexrhogr}), the angular scaling is not a useful procedure for the derivation of \g\, and a pinning--free quantity must be considered in order to get the true intrinsic anisotropy \g.
\begin{figure}[ht]
\includegraphics[width=8 cm]{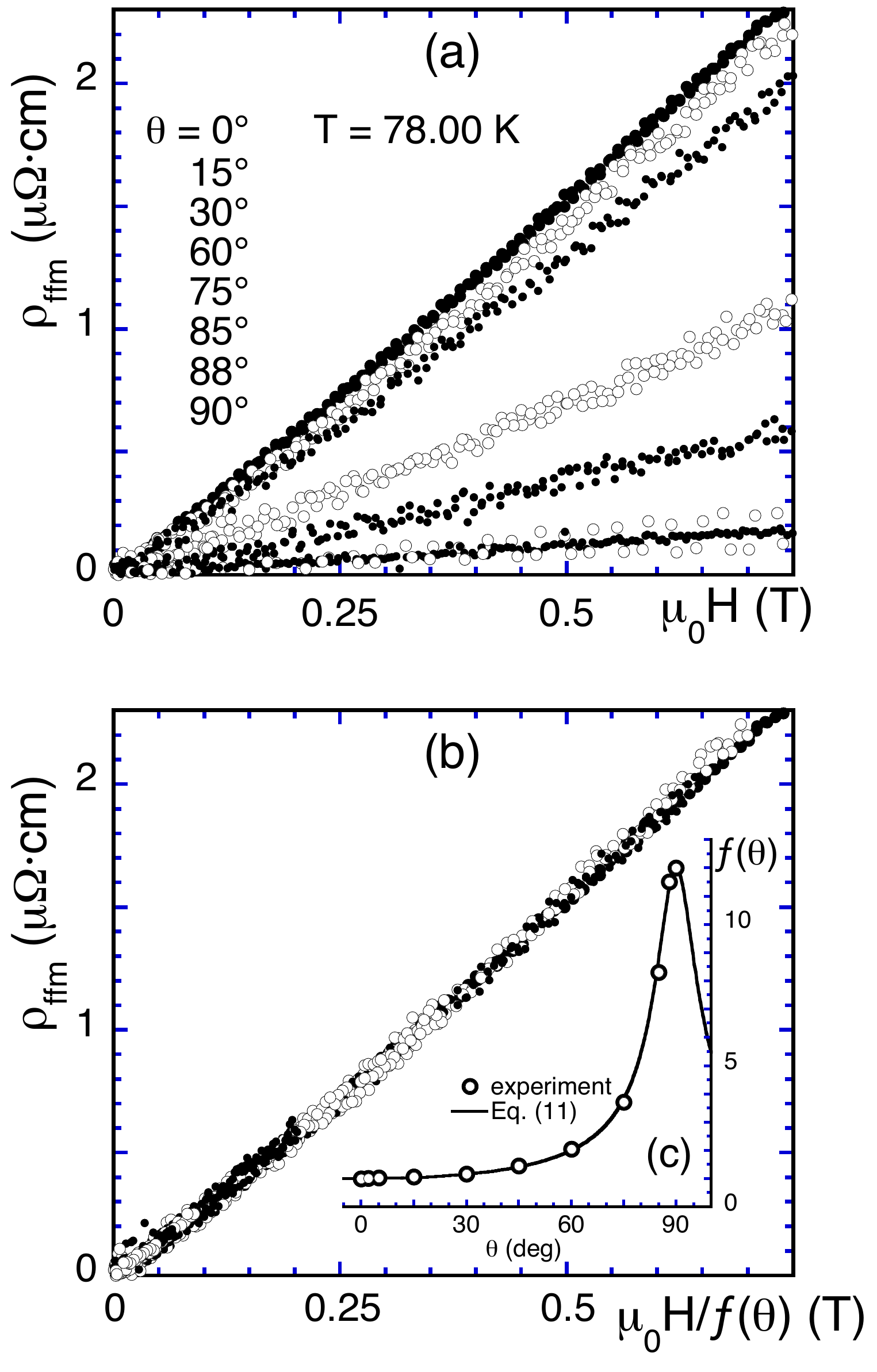}
\caption{Demonstration of the successful angular scaling for the flux--flow resistivity derived from the data in Fig. \ref{fig:noscaling} (only 30\% of data or less is shown to avoid crowding), Full and open symbols are alternated to ease reading of the scaled data. (a) experimental flux--flow resistivity $\rho_{\ff m}$ as a function of the applied magnetic field $H$ at selected angles $\theta$. (b) Scaling  $\rho_{\ff m}(H/f(\theta))$ of the data in (a). Inset to (b): experimental scaling function $f(\theta)$ and fit to Eq.\ref{eq:f}, yielding \g = 6.}
\label{fig:scalingrhoff}
\end{figure}
We then extract, for the same set of data, the (pinning--unrelated) flux--flow resistivity $\rho_{\ff m}(H,\theta)$ (Fig. \ref{fig:scalingrhoff}a) and we look for the angular scaling, which is indeed possible as depicted in Fig.\ref{fig:scalingrhoff}b. Thus, the angular scaling takes place for the flux--flow resistivity, and it is then possible to derive an experimental scaling function $f(\theta)$ defined as the scaling factor such that by plotting the curves for $\rho_{\ff m}$ vs. $H/f(\theta)$, a scaling is accomplished. It is then possible to compare $f(\theta)$ to the theoretical expectations to obtain \g\,, but it should be reminded that the geometry plays a crucial role, as described in Sec.\ref{sec:micro}. We consider the case where $\rho_{\ff m}(H) \propto H^{\beta}$. The case $\beta=1$ is the behaviour predicted by many models for the free--flow of vortices \cite{Tinkham1964, Bardeen1965, Gor'kov1976}, and exhibited by our data. In this case, from (i) Eq.\ref{eq:rhoff}, that states the difference between the measured $\rho_{\ff m}$ and the $ab$ plane component of the flux--flow resistivity tensor, (ii) the scaling evidence that $\rho_{\ff m}(H,\theta) =\rho_{\ff m}(H/f(\theta))$, and (iii) Eq. \ref{eq:fteta}, one gets the expected form for the scaling function:
\begin{equation}
\label{eq:f}
    f(\theta)=\epsilon^{-1}(\theta)\left[f_L(\theta)\right]^\frac{1}{\beta}=
    \left(\frac{1}{\gamma^{-2}\sin^2\theta+\cos^2\theta}\right)^\frac{1}{2}
    \times
    \left[\frac{\gamma^{-2}\sin^2\theta+\cos^2\theta}{\frac{\gamma^{-2}}{2}\sin^2\theta+\cos^2\theta}\right]^\frac{1}{\beta}
\end{equation}
With respect to the conventional scaling factor \eps,  the varying Lorentz force contributes with the second term in square brackets. Remarkably, there is still only \g\, as the single parameter in the scaling function. It is useful to remark that in the present case $\beta=1$, for $\theta=90^{\circ}$ the scaling factor is not \g\, but 2\g, due to the halving of the Lorentz force (see the geometry in Fig.\ref{fig:geometry}).

The experimental scaling function $f(\theta)$ for the data in Figs. \ref{fig:noscaling},\ref{fig:scalingrhoff} is reported in Fig.\ref{fig:scalingrhoff}c, together with the fit of Eq.\ref{eq:f} which yields \g=6.
\begin{figure}[ht]
\includegraphics[width= 7.5cm]{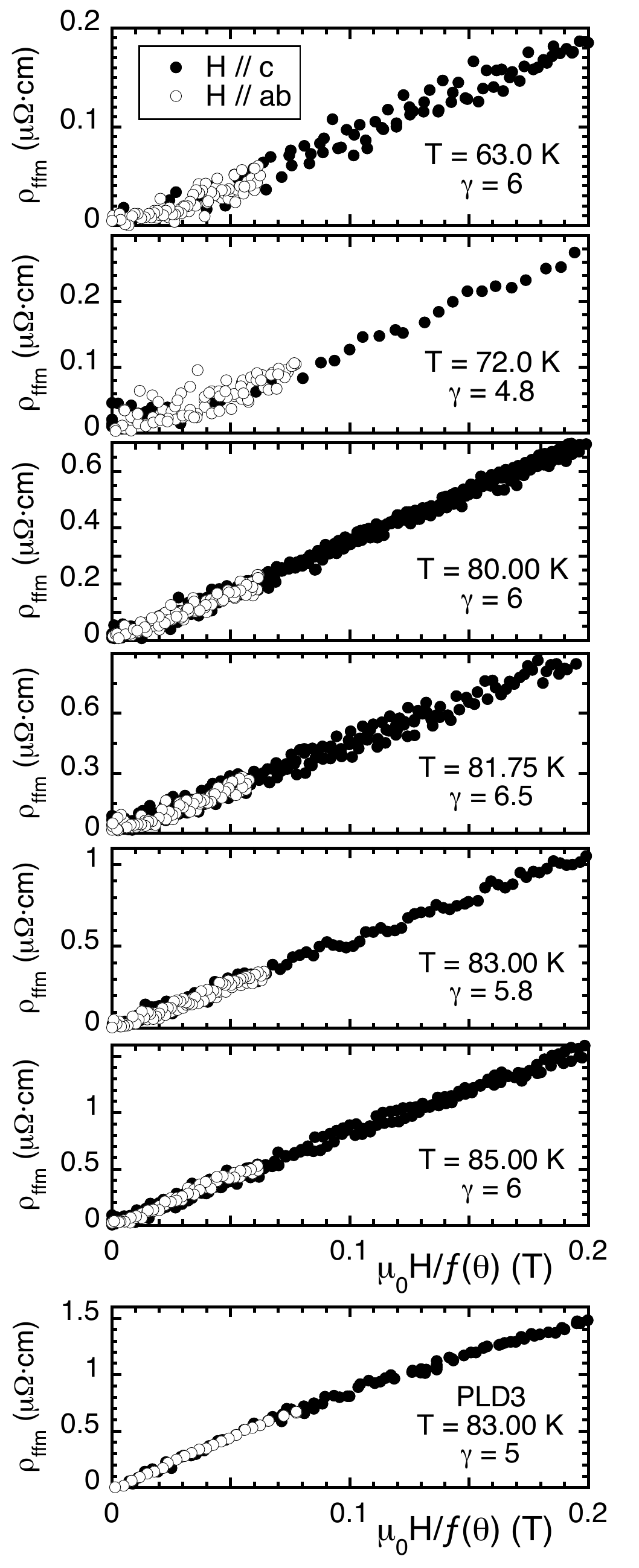}
\caption{Simplified angular scaling for the flux--flow resistivity in samples CSD3 and PLD3 (bottom panel): the curve measured at $\theta=90^{\circ}$ is scaled onto the curve taken in perpendicular ($\theta=0^{\circ}$) orientation. Only the low field range is shown to appreciate the scaling. For the data at $\theta=0^{\circ}$, only 10\% of data is plotted to avoid crowding.}
\label{fig:littlescaling}
\end{figure}
\begin{figure}[ht]
\includegraphics[width= 7cm]{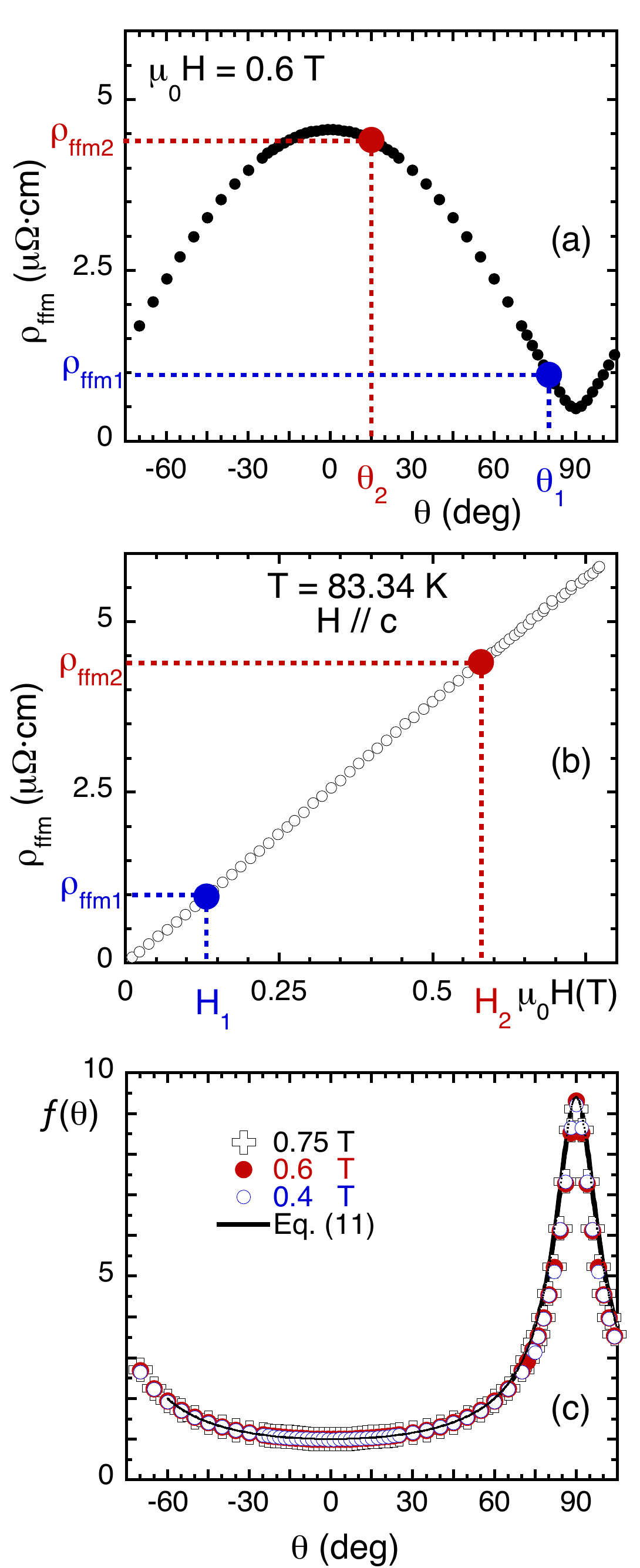}
\caption{Collapsing of the angular rotations for the flux--flow resistivity, \mbox{$\rho_{\ff m}(H=0.6\;\rm{T}, \theta)$} (a), over the field--dependent curve  $\rho_{\ff m}(H, \theta=0)$ (b), and obtained $f(\theta)$ at three different fields yielding \g=4.7 (c). Data taken in sample CSD1 at $T=\;$83.34 K and $\mu_0H=\;$0.6 T. The procedure is graphically exemplified for two points (red and blue).}
\label{fig:rotationscaling}
\end{figure}

Once the scaling of $\rho_{ff m}$ has been demonstrated, it is possible to obtain $f(\theta)$ and \g\, from less time--consuming measurements. To evaluate only \g, a first possibility is to simply measure $\rho_{\ff m}$ for $\theta=0^{\circ}$ and $\theta=90^{\circ}$. The scaling of $\rho_{ff m}(H, 90^{\circ})$ over $\rho_{ff m}(H, 0^{\circ})$ directly yields \g, although the shape of $f(\theta)$ cannot be known. This is demonstrated in Fig. \ref{fig:littlescaling}, where this procedure is performed at different temperatures in samples CSD3 and at one temperature in sample PLD3. We note that the measured signal in parallel orientation is weak, leading to a large uncertainty ($\pm0.5$) in such estimates of \g. 

Another way to obtain \g\, is to rely on the angular rotations: $\rho_{ff m}$ measurements taken at fixed field $H$ and varying angle. This procedure has been explained in detail in \cite{Pompeo2014b,Bartolome2019}, and an example is reported in Fig. \ref{fig:rotationscaling}. In this procedure, each datum point of the rotation is reported on the perpendicular field curve by rescaling its field as $H/f_{exp}(\theta)$. By forcing the scaling, one gets the experimental $f_{exp}(\theta)$. Clearly, this is possible only if the existence of the scaling has been previously assessed. A successful fitting of $f_{exp}(\theta)$ to Eq. \ref{eq:f} yields \g. 

We now discuss the results for the scaling function $f(\theta)$ and for the resulting \g\, in the full body of our samples. Table I  reports the values of \g\, that we obtained on different samples, at different temperatures and different fields. There is no particular trend depending on the sample type. Moreover, it is noteworthy that in all cases we obtain a similar intrinsic anisotropy factor \g. By assuming that the interval covered by our data represents a uniform distribution for \g\, (a worst case), we obtain $\gamma \pm \sigma_{\gamma} = 5.3\pm 0.7$, where $\sigma_{\gamma}$ is the standard deviation for a uniform distribution. This is remarkable due to the very different nature of the samples: CSD pristine, CSD with very different nanostrain, PLD with nanorods. This is a central result of this paper: although nanostructuring is very different, the intrinsic anisotropy \g\, is the same. We should note that the values that we consistently obtain for \g\, are  lower than the reported values $\gamma \sim 7$ for pure, single--crystalline YBCO \cite{Roulin1996,Naughton1988,Kobayashi1995} and at the lower edge of the reported values $\gamma \sim 5-7$ in epitaxial pristine thin films \cite{Sarti1997a,Sarti1997b,Xu2015b,Llordes2012,Miura2015}. Although we do not have an explanation for this result, the very different nature of our samples would point to some general effect in the electronic structure itself toward a lesser anisotropy. Normal--state measurements of the mass tensor would be useful to assess this aspect.
\Table{
\label{tabl3}
Collection of all measured values for \g.
}
\br
&Measuring&Temperature&Intrinsic anisotropy factor$^{\rm b}$\\
Sample&method$^{\rm a}$&range& $\gamma$\\
\mr
CSD1 & P & 77.70 K -- 83.34 K & 4.8 $\pm$ 0.2\\
 & R & 77.70 K -- 83.34 K & 5.3 $\pm$ 0.7\\
CSD2$^c$ & R & 77.70 K & 4.5 $\pm$ 0.5\\
CSD3 & S & 78.00 K & 6.0 $\pm$ 0.2\\
 & P & 63.00 K -- 85.00 K & 5.7 $\pm$ 0.8\\
PLD1 & S & 81.00 K & 5.8 $\pm$ 0.2\\
PLD2$^d$ & R & 80.00 K & 5.0 $\pm$ 0.2\\
PLD3 & P & 83.00 K & 5.0 $\pm$ 0.2\\
\br
\end{tabular}
\item[] $^{\rm a}$ S: full angular scaling, as in Fig.\ref{fig:scalingrhoff}. P: scaling of the parallel curves over perpendicular, as in Fig.\ref{fig:littlescaling}. R: scaling of the rotations at fixed fields, as in Fig.\ref{fig:rotationscaling}.
\item[] $^{\rm b}$ The uncertainty is evaluated as the half width of the dispersion of the measured anisotropies for the case of many sets of data (maximum uncertainty), and as the deviation from an appreciable scaling when only one measurement exists.
\item[] $^{\rm c}$ Ref. \cite{Bartolome2019}.
\item[] $^{\rm d}$ Ref. \cite{Pompeo2013}.
\end{indented}
\end{table}

\section{Anisotropic pinning}
\label{sec:pinning}
Having settled the value of the anisotropy factor in the previous Section, we turn now to the discussion of the effect of directional pinning on our microwave data. We then focus on the angular dependence of the depinning frequency, $\nu_p(\theta)$. According to Sec.\ref{sec:micro}, the angular dependence of $\nu_p$ is not affected by the Lorentz force contribution \cite{Pompeo2018a}. The pinning anisotropy should then come entirely from the field orientation. We discuss briefly this point. Since we are dealing now with a pinning--related property, we expect an angular dependence of $\nu_p$ which is an admixture of (i) the scaled field $H\epsilon(\theta)$ and (ii) the anisotropic pinning, which depends on any possible directional pinning and distortion of the intrinsic anisotropy due to, e.g., 3D defects. Thus, no universal behaviour can be expected in superconductors with so large differences in nanostructuring like our samples. This Section will then overview the main features that can emerge from the analysis of the angular dependent microwave surface impedance, underlying the potential of the high frequency method, without attempting at a full explanation of all the features observed in $\nu_p$.
\begin{figure}[ht]
\includegraphics[width= 8cm]{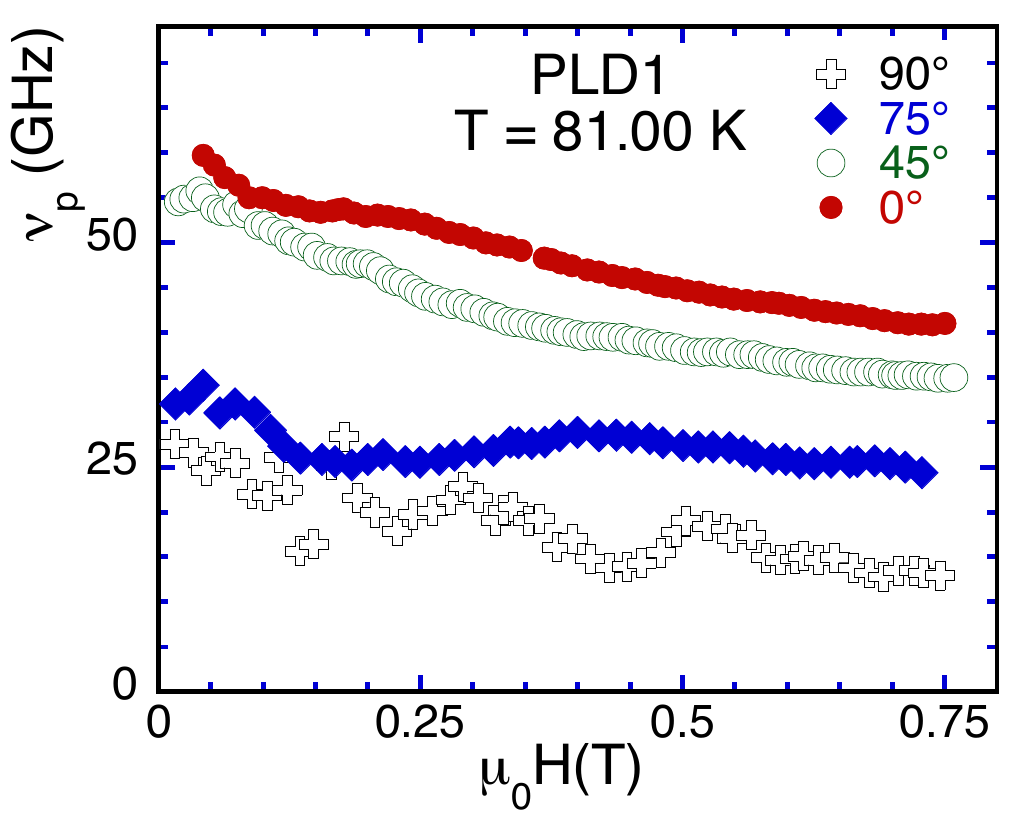}
\caption{Depinning frequency $\nu_p$ in sample PLD1 as a function of the field and different field orientations, for $T=\;$81 K. It is immediately seen that no field scaling $H/f(\theta)$ is possible. Moreover, the depinning frequency has an {\em inverse} angular dependence with respect to the expectations, because of the strong $c$--axis pinning due to BZO nanorods. This figure should be compared to Fig.\ref{fig:scalingrhoff}a of $\rho_{ff m}(H)$ at various angles.}
\label{fig:nupPLD1}
\end{figure}
\begin{figure}[ht]
\includegraphics[width= 8cm]{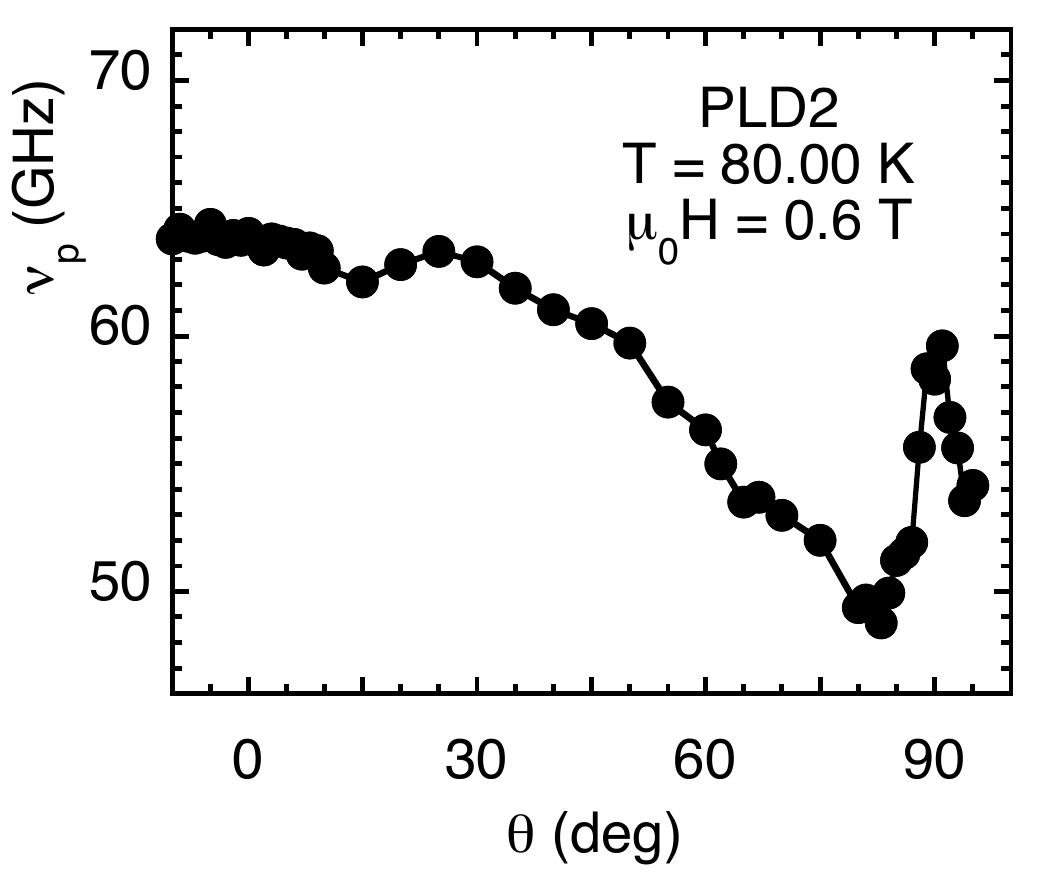}
\caption{Depinning frequency $\nu_p$ in sample PLD2 as a function of the field orientation at the fixed field $\mu_0H=\;$0.6 T, and $T=\;$80 K. The broad maximum at $\theta=0^{\circ}$ is due to BZO nanorods aligned with the $c$--axis with a little splay, the peak at $\theta=90^{\circ}$ originates from ab--plane pinning. This figure should be compared to Fig.\ref{fig:rotationscaling} of $\rho_{ff m}(\theta)$ at fixed magnetic field.}
\label{fig:nupPLD2}
\end{figure}

We first remark that the angular scaling does not in general apply to pinning--affected properties, and thus it is not a general feature. To this aim, we report in Figure~\ref{fig:nupPLD1} a subset of the curves of $\nu_p$ vs. $H$ at different angles in sample PLD1. This figure demonstrates that the scaling does not take place in $\nu_p$: the field dependence is basically flat, but the vertical scale is much different, so no scaling of the curves can exist. Moreover, the curves exhibit an {\em inverse} dependence of the pinning property with the field orientation: in a pure material such as a single crystal $\nu_p$ has a maximum for $\theta=90^{\circ}$ \cite{Golosovsky1996}, due to the ab--plane pinning, whereas here $\nu_p$ has a maximum for $\theta=0^{\circ}$. This is not surprising when the microstructure of the PLD sample is taken into account: the BZO nanorods are oriented with little splay \cite{Augieri2010}, so strong c--axis pinning is expected. Thus, the angular scaling cannot be applied to the pinning properties.

In fact, the angular dependence of the depinning frequency can be much more complicated \cite{Pompeo2014b}. Figure \ref{fig:nupPLD2} reports the full angular dependence of $\nu_p$ at a fixed field $\mu_0H=0.6\;$T in sample PLD2 at $T=\;$80 K. We first note that the absolute value of $\nu_p$ is rather different than in sample PLD1: this is a manifestation of the strongly sample--dependent pinning properties. The full angular scan reveals the effect of the directional pinning: a broad peak exists at $\theta=0^{\circ}$, as a consequence of the BZO nanorods, twin planes and grain boundaries, and a sharp peak develops around $\theta=90^{\circ}$, where $ab$--plane pinning becomes effective. This is fully consistent with the previous findings in PLD samples. There, comparison between $J_c$ and microwave data showed \cite{Pompeo2013} that the BZO nanorods gave a much different contribution in dc and at high frequencies. In particular, it was argued that the main effect of nanorods was to pin very strongly a reduced number of vortices that acted then as a caging structure for the remaining flux lines. This effect was much less effective at microwaves, where tiny vortex oscillations are involved. Thus, while $J_c$ presented a sharp peak at $\theta=0^{\circ}$, microwaves reported only a broad maximum. 
\begin{figure}[ht]
\includegraphics[width= 8cm]{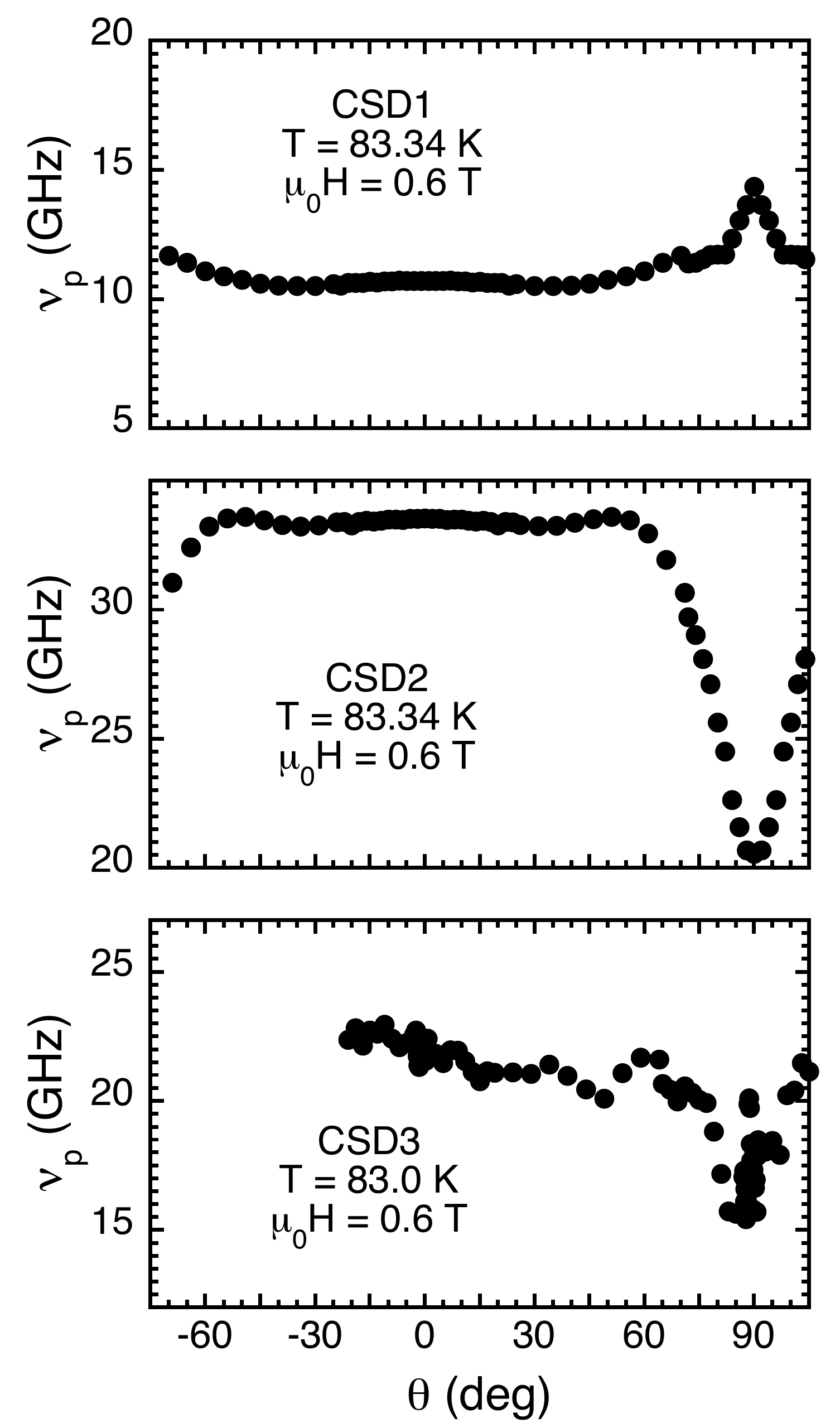}\\
\caption{Depinning frequency $\nu_p$ in CSD samples as a function of the field orientation at the fixed field $\mu_0H=\;$0.6 T, and $T=\;$80 K. The CSD1 sample (pristine) exhibits a behaviour similar to PLD1: a sharp peak at $\theta=90^{\circ}$ due to intrinsic (ab--planes) pinning, and a broad maximum at $\theta=0^{\circ}$ due to twin planes or grain boundaries. Nanoparticles in samples CSD2 and CSD3 make pinning much more isotropic: the anisotropy of $\rho_{ff m}$ prevails over the anisotropy of the pinning constant, and as a consequence a dip appears approaching $\theta=90^{\circ}$ (see also text). In CSD3 a remaining of the ab--plane pinning exists.}
\label{fig:nupCSD}
\end{figure}

The behaviour of the CSD samples is particularly interesting. We remind that in these samples the angular dependence of the dc critical current density exhibits a sample--dependent reduced effective anisotropy \cite{Bartolome2019}, evaluated from pinning--affected properties, whereas the intrinsic anisotropy \g$\simeq$5.3 determined in Sec.\ref{sec:scaling} did not differ significantly from sample to sample. Figure \ref{fig:nupCSD} presents the angular scans $\nu_p(\theta)$ in samples CSD1 (pristine), CSD2 and CSD3, at the same temperature $\simeq$83 K and field $\mu_0H=$0.6 T. The pristine sample behaves similarly to the PLD2 sample: a broad maximum in $\nu_p$ at $\theta=0^{\circ}$, and a peak at $\theta=90^{\circ}$.  The absence of nanorods makes the maximum at $\theta=0^{\circ}$ much less evident than in sample PLD2, and much lower than the peak at $\theta=90^{\circ}$, but still the unavoidable presence of twin planes and grain boundaries determines an enhanced pinning when $H \parallel c$.

When BYTO nanoparticles are added, the entire behaviour of $\nu_p(\theta)$ flattens out, apart from a {\em minimum} at $\theta=0^{\circ}$ (some vestiges of twin planes or grain boundaries pinning are still detectable as a very weak maximum at $\theta=0^{\circ}$, Fig.\ref{fig:nupCSD}b). Again, this is fully consistent with the nanostructure and with the effect on $J_c(\theta)$. We remind that $\nu_p=k_p/2\pi\eta$, and that $\eta(\theta)=\Phi_0\mu_0H/\rho_{\ff}(\theta)$ is affected by intrinsic anisotropy only, as shown in Sec.\ref{sec:scaling}. Since $\nu_p\propto k_p(\theta)\rho_{ff}(\theta)$, the anisotropy in $	\nu_p$ comes from a competition between the anisotropies of $k_p$ and $\rho_{ff}$, where the anisotropy of $\rho_{ff}$ is the intrinsic anisotropy. In crystalline materials \cite{Golosovsky1996}, $k_p$ exhibits a very large maximum at $\theta=90^{\circ}$ due to $ab$ planes intrinsic pinning (see Sec.\ref{sec:intro}, footnote), and as a consequence its anisotropy prevails and $\nu_p$ exhibits a peak at $\theta=0^{\circ}$. This is the behaviour exhibited by sample CSD1 in Fig.\ref{fig:nupCSD}a. Since the intrinsic anisotropy \g\, is the same for all samples, so is the angular dependence of $\eta$ (or of $\rho_{ff})$. Thus, the results shown in Figures~\ref{fig:nupCSD}b,~\ref{fig:nupCSD}c clearly show that the introduction of nanoparticles leads to a dramatic flattening of $k_p$: a strong reduction of the pinning anisotropy. Since the intrinsic anisotropy does not change, the anisotropy of $\rho_{ff}$ does not change and then a minimum in $\nu_p$ arises.
We note that the temperature of our experiment is rather close to the transition, but the intrinsic pinning effects are still prominent, at least in some samples. In fact, although the divergence of the out--of--plane coherence length $\xi_{\perp}$ tends to weaken the effect of the layered structure on pinning, the modulation of the potential has been observed to persist up to 0.5 K below the critical temperature \cite{Kwok1991}.

A complete treatment of the angular dependence of pinning as probed at microwave frequencies would require a much lengthier analysis. However, from the sample  measurements here presented there are several points that can be concluded. First, directional pinning manifests itself even at microwave frequencies, with very small vortex displacement. Following the same physical reason, nanorods are not as effective as in dc: no sharp peak is observed when the field is aligned with nanorods, contrary to measurements of $J_c$ \cite{Civale2004,Foltyn2007,Pompeo2013}. Second, the effect of nanoparticles in flattening the angular dependence of pinning is exceptionally evident in our CSD samples: the anisotropy is reversed in the depinning frequency, meaning that the pinning constant $k_p$ has a very reduced anisotropy, and certainly less than the intrinsic anisotropy exhibited by the flux--flow resistivity.

\section{Conclusions}
\label{sec:conc}
We have presented a broad analysis of experimental results for the angular dependence of the microwave surface impedance in PLD and CSD samples, with different kind of nanostructuring, with the aim of discriminating the intrinsic anisotropy from the pinning anisotropy, giving the opportunity to study them separately. The measured complex resistivity allowed to extract the flux--flow resistivity and the depinning constant. The flux--flow resistivity, being unrelated to pinning, exhibited perfect angular scaling properties according to the BGL scaling rule. From the scaling, we obtained \g=5.3$\pm$0.7 in all our samples, irrespective of the kind of defects. By contrast, the depinning frequency $\nu_p$ showed an angular dependence strongly affected by directional pinning. In particular, in PLD and pristine CSD the contribution of $ab$--planes and $c$--axis directional pinning (nanorods, twin planes, grain boundaries) is evident. Instead, in CSD samples with nanoparticles, that lead to increased nanostrain with respect to pristine samples, the pinning anisotropy was reduced well below the intrinsic anisotropy. The microwave measurements are then demonstrated to be a powerful tool to ascertain the pinning and intrinsic anisotropy separately.

\ack
This work has been carried out within the framework of the EUROfusion Consortium and has received funding from the Euratom Research and Training Programme 2014-2018 and 2019-2020 under grant agreement No 633053. The views and opinions expressed herein do not necessarily reflect those of the European Commission. We thank Benedetta Belli and Raffaella Rogai for help in data acquisition. We acknowledge financial support from Spanish Ministry of Economy and Competitiveness through the ÒSevero OchoaÓ Programme for Centres of Excellence in R\&D (SEV-2015-0496), SUMATE project RTI2018-095853-B-C21, co-financed by the European Regional Development Fund, COST-Nanohybri CA16218, DWARFS project (MAT2017-83468-R) and Catalan Government with 2017-SGR-1519 and XRE4S. \\

\vspace{1cm}

\providecommand{\newblock}{}

\end{document}